\documentstyle[preprint,tighten,aps,psfig]{revtex}
\begin{document}
\title{Comment on ``Chiral Corrections in Hadron Spectroscopy''}
\author{ L. Ya. Glozman}
\address{ 
 High Energy Accelerator Research Organization (KEK),
Tanashi Branch, Tanashi, Tokyo 188-8501, Japan}
\maketitle

\begin{abstract} 
It is shown that the principal pattern in baryon spectroscopy,
which is associated with the flavor-spin dependent hyperfine interaction,
is due to the spontaneous breaking of chiral symmetry in QCD and thus
cannot be addressed by chiral perturbation theory,
which is based on the explicit chiral symmetry breaking.
\end{abstract}

\bigskip
\bigskip

In a recent preprint \cite{THOMAS} Thomas and Krein
questioned the foundations of the description of the 
baryon spectrum with
a chiral constituent quark model \cite{GR,GPVW}. The 
argument made was that the splitting pattern implied
by the operator

\begin{equation}
-\vec{ \tau}_{i} 
\cdot \vec{\tau}_{j}
\vec{\sigma}_i \cdot \vec{\sigma}_j,
\label{GBE1}
\end{equation}

\noindent 
should be inconsistent with chiral symmetry because it is
inconsistent with the leading nonanalytic contribution
to baryon mass predicted by the chiral perturbation theory (ChPT).
I here show that ChPT has no bearing on this issue.\\

There are two distinct aspects of chiral
symmetry in QCD. The first one is the spontaneous (dynamical)
breaking of chiral symmetry, which leads to the 
constituent (dynamical) masses
of quarks (which are related to the quark condensates), massless Goldstone
bosons, and their couplings to constituent quarks. 
As a result there appears in
the chiral limit the massless Goldstone boson exchange
interaction between the constituent quarks:

\begin{equation}
V_\chi = - \frac{g^2}{4\pi} \frac{1}{12m^2}
\vec{ \tau}_{i} 
\cdot \vec{\tau}_{j}
\vec{\sigma}_i \cdot \vec{\sigma}_j 4\pi \delta(\vec r),
\label{GBE2}
\end{equation}

\noindent
where the tensor force component which is irrelevant
to discussion here, has been dropped.

In reality the contact interaction (\ref{GBE2}) is smeared 
by the finite size of constituent quarks and pions
\footnote{Note that in the chiral limit the volume integral
constraint, $\int d \vec r V_\pi(\vec r)=0$, does not apply
\cite{GLOZ} as in this case the pion Green function 
$\sim -\frac{1}{(\vec q)^2}$ exactly cancels the $\sim (\vec q)^2$
behaviour supplied by the pion-quark vertices.}.
Following to Thomas and Krein I consider for simplicity only
the pion-exchange part of the complete Goldstone boson exchange
interaction.\\

The second aspect of the chiral symmetry is its explicit breaking
by the nonzero mass of current quarks. This will slightly modify
the constituent mass $m$ because of the direct contribution of current
quark mass and because of pion loop self-energy corrections,
which are subject to renormalization. The other implication of
the explicit chiral symmetry breaking is that the pion obtains
a finite mass $\mu$, which to leading order is illustrated by the current
algebra (Gell-Mann - Oakes - Renner relations). As a consequence
there appears a long-range Yukawa potential interaction between
quarks

\begin{equation}
V_\chi =  \frac{g^2}{4\pi} \frac{1}{12m^2}
\vec{ \tau}_{i} 
\cdot \vec{\tau}_{j}
\vec{\sigma}_i \cdot \vec{\sigma}_j
\left (\mu^2\frac{e^{-\mu r}}{r} - 4\pi \delta(\vec r)\right ).
\label{GBE3}
\end{equation}

\noindent
Chiral perturbation theory only concerns the
implications of the explicit chiral symmetry breaking term in
(\ref{GBE3}) and hence by
definition cannot be used to derive the expression (\ref{GBE2}).\\

The idea of the chiral constituent quark model \cite{GR,GPVW,GLOZ}
is that the main features of the baryon spectrum are
supplied by the spontaneous breaking of chiral symmetry, i.e.
 by the constituent mass of quarks and the interaction (\ref{GBE2})
between them\footnote{ Note that the short-range interaction
of the same type comes also from the $\rho$-meson-exchange and/or
correlated two-pion-exchange, etc, and there are important reasons
to believe that the latter contributions are also important \cite{GLOZ}.}. 
As a consequence the $N$ and $\Delta$ are split already in the chiral 
limit, as it must be. The expressions (in the notation of ref. \cite{GR})

$$ M_N = M_0 - 15 P_{00}^{\pi},$$
\begin{equation}
M_\Delta = M_0 - 3P_{00}^{\pi} \label{D},
\end{equation}

\noindent
where $P_{00}^{\pi}$ is positive,  arise from the interaction
(\ref{GBE2}). The long-range Yukawa tail, which
has the opposite sign represents only a small perturbation. 
It is in fact possible to obtain a near perfect fit of the 
baryon spectrum
in a dynamical 3-body calculation in the chiral limit, neglecting 
the long-range Yukawa tail contribution, with a quality
even better than that of \cite{GPVW}.\\

The implication is that ChPT has no bearing on the interactions
(\ref{GBE1}) and (\ref{GBE2}) (nor on the
the expressions (\ref{D}), which should be considered
as leading order contributions ($\sim m_c^0$, where $m_c$ is current
quark mass)
within chiral perturbation
theory. This does not mean, however, that the systematic corrections
from the finite meson (current quark) mass should be ignored. \\

The pion-exchange Yukawa tail contribution contains at the
same time part of corrections from the leading nonanalytic
order, $\mu ^3$, as well as part of the corrections
from the other orders, with opposite sign. 
A rough idea about importance of these corrections for the
$N$ and $\Delta$ can be obtained from the comparison
of the contributions of the first and second terms in (\ref{GBE3})
in  nonperturbative calculations \cite{GPVW}.
The former one turns out to be much smaller than the latter.
This is because of a small matter radius of the $N$ and $\Delta$
\cite{GV}. For highly excited states, however, the role of the
Yukawa tail increases because of a bigger baryon size and thus
the importance of the ChPT corrections should be expected to 
increase.
To
consider these corrections systematically one definitely needs
to consider the  loop contributions to the interactions
between constituent quarks as well as the couplings to
decay channels, which is rather involved task. This  task is one
for constituent
quark chiral perturbation theory which is awaiting 
practical implementation.

\end{document}